\documentclass[letterpaper]{sfchem}
\usepackage{graphicx}

\begin{document}

\title{Chemistry and Dynamics in Pre-Protostellar Cores}

\author{Jeong-Eun\ Lee\inst{1} \and Neal J. Evans II\inst{1} \and Yancy L.
Shirley\inst{1} \and Ken'ichi Tatematsu\inst{2}}  
  \institute{Department of Astronomy, The University of Texas at Austin, 
  1 University Station C1400, Austin, Texas 78712--0259 \and   
  National Astronomical Observatory of Japan, 2-21-1 Osawa, Mitaka,
  Tokyo 181-8588, Japan
  } 
\authorrunning{Lee, Evans, Shirley, and Tatemtsu}
\titlerunning{Chemistry and Dynamics in Pre-Protostellar Cores}

\maketitle 

\begin{abstract}

We have compared the intensity distribution of molecular line emission with that
of dust continuum emission, and modeled molecular line profiles 
in three different preprotostellar cores in order to test 
how dynamical evolution is related to chemical evolution, and whether we 
can use different chemical tracers to identify specific dynamical evolutionary 
stages. We used dust continuum emission to obtain the input 
density and temperature structures by calculating radiative transfer of dust 
emission.   
Our results show that chemical evolution is dependent on dynamical processes,
which can give different evolutionary timescales, as well as the density 
structure of the core. 


\end{abstract}

\section{Introduction and Observations}

Low mass star formation has received more detailed study, both in theory and 
observation, compared to massive star formation. An evolutionary seqence of low
mass star formation based on the spectral energy distribution (SED) has been
developed (Adams et al. 1987). However, the initial conditions of collapse are 
still poorly understood. 
Preprotostellar cores (PPCs) are believed to be gravitational bound, but they 
have no central hydrostatic protostar (Ward-Thompson et al. 1994).
Therefore, PPCs can be considered as the potential sites of future star 
formation and give us chances to probe the initial conditions of star formation.
Evans et al. (2001) modeled dust continuum in three PPCs using Bonnor-Ebert 
spheres and showed that the dust temperature decreases toward the 
center because PPCs are heated only by the interstellar radiation field, 
and the radiation from the cores is optically thin. 
Therefore, they argued that the 
actual density structure of a preprotostellar core is more centrally condensed 
compared to the intensity distribution of the dust continuum emission. 

Several studies on chemical 
evolution in PPCs have been done by Bergin \& Langer (1997), Aikawa et al. 
(2001), Caselli et al. (2002), and Li et al. (2002). 
They tested different dynamical scenarios, but all included the interaction
of gas phase molecules with the surfaces of dust grains
as well as gas phase chemistry and showed that some molecules such as 
CS and CO are depleted substantially from the gas phase as the density 
increases. 
Those molecules have high binding energy onto 
the surfaces of dust grains, so their timescales for evaporation by thermal 
energy or cosmic rays are much longer than those of molecules that 
have lower binding energy. For instance, nitrogen bearing molecules such as 
N$_2$H$^+$ and NH$_3$, which become abundant in later stages of dynamical 
evolution and have low binding energy, do not show as significant depletion
as CO and CS. Comparison of the abundance change of a molecule with time and 
the differential distributions of molecules in a given time seems to be able to 
trace the dynamical evolutionary 
stages. We tested this idea by comparing the line emission from several 
molecules
that have different chemical evolution timescales with dust emission,
which can trace physical conditions,
in three PPCs: L1512, L1544, and L1689B. 
 
We observed the three PPCs in C$^{18}$O ($\rm J=2-1$ and $\rm J=3-2$), 
C$^{17}$O ($\rm J=2-1$), DCO$^+$ ($\rm J=3-2$), $\rm HCO^+$ ($\rm J=3-2$), and
$\rm H^{13}CO^{+}$ ($\rm J=3-2$) with the 10.4 m telescope of the Caltech 
Submillimeter Observatory (CSO) at Mauna Kea, Hawaii from 1995 to 2002.
We also observed these cores in $\rm H^{13}CO^{+}$ ($\rm J=1-0$), $\rm N_2H^+$
($\rm J=1-0$), and CCS ($\rm N_J=4_3-3_2$) with the 45 m telescope
of the Nobeyama Radio Observatory in Japan on January 2002.
See Lee et al. (2003) for details about observations.

\section{Simple Analysis}

We compare, first, the H$_2$ column densities calculated from molecular lines 
(C$^{18}$O $\rm J=2-1$ and C$^{17}$O $\rm J=2-1$) and dust continuum,
which can be used as the tracer of column density least affected by chemistry
(Figure 1.a). 
We assume that temperature and abundance are constant along the line of sight,
that gas and dust have the same temperature, that all levels are in LTE, 
and that lines are optically thin. The H$_2$ column densities 
calculated from C$^{18}$O $\rm J=2-1$ are smaller than those calculated from
850 $\mu$m dust emission by a factor of 10 to 25. 
Also, the C$^{18}$O emission shows no peak at the position of the peak of
dust continuum emission. If this line is optically 
thin, so that it traces all material along the line of sight, the difference 
would be caused by the depletion of CO.
The optical depth of C$^{17}$O $\rm J=2-1$ can be calculated by fitting the
relative strength of its hyperfine components, and the calculated optical
depths in three PPCs show that the optical depth of C$^{17}$O $\rm J=2-1$
is negligible in all three cores, but C$^{18}$O $\rm J=2-1$ is optically thick 
in L1544 and L1689B (see \S 4.2.1 of Lee et al. 2003).   
Therefore, if we use an optically thick line, the depletion factor can be 
overestimated. The comparison between the H$_2$ column densities calculated from
C$^{17}$O $\rm J=2-1$ and 850 $\mu$m dust emission shows that the CO molecule
is significantly depleted in L1512 and L1544 but not in L1689B.  

We also compare the distributions of the intensity of molecular emission 
and dust continuum emission (Figure 1.b). 
L1512 is the weakest, and L1544 and L1689B are similar in dust continuum. 
However, the CCS intensity of L1689B is more similar to that of L1512. 
Even though the distribution of CCS is very different from that of the dust 
continuum, it does not show the central dip expected if the
molecule is depleted significantly as some chemical modelings show.   
We suggest that this trend would be caused by the significant depletion of CO, 
helping C$^+$ to react more with molecules that are not oxygen-bearing species
(Li et al. 2002). 
The intensity of N$_2$H$^+$ follows dust emission well in L1544 and L1689B, 
but it is weaker in L1689B than in L1544. 
The difference is possible because CO, which is a destroyer of N$_2$H$^+$, is
much more depleted in L1544.
On the other hand, in the central region of L1512, the N$_2$H$^+$ distribution 
is flatter than the dust emission indicating the possible depletion of 
N$_2$H$^+$. 

\begin{figure}[ht]
\centering
\includegraphics[height=0.8\linewidth]{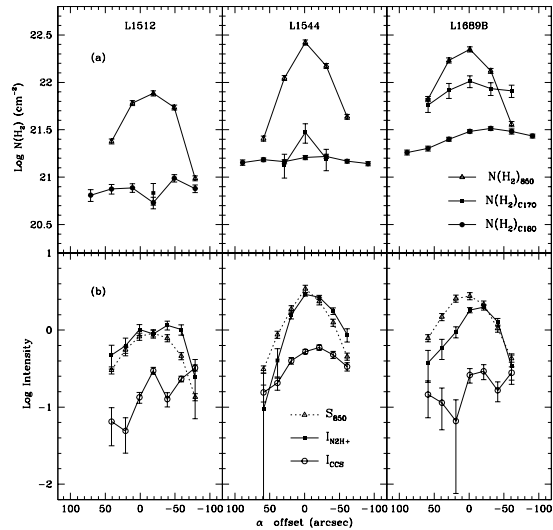}
\caption{
(a) compares the H$_2$ column densities 
calculated from the 850 $\rm \mu m$ dust continuum emission,
$\rm C^{18}O$ $\rm J=2-1$ (without the correction of $\tau$), and
$\rm C^{17}O$ $\rm J=2-1$ (with correction of $\tau$)
through the cuts marked in Figure 2 of Lee et al. (2003).
(b) compares the integrated intensities of CCS
$\rm N_J=4_3-3_2$ and $\rm N_2H^+$ $\rm J=1-0$
lines with $S_{850}$ through the same cuts in (a).
Here, $S_{850}$ is shifted by 0.9.
\label{fig-single}}
\end{figure}

\section{Detailed Models}

The simple method used in the previous section has limitations for 
quantitative analysis because it cannot include the variation of the excitation
temperature and the abundance of a molecule along the line of sight. 
Therefore, we used the Monte Carlo (MC) code and the virtual telescope 
simulation code developed by Choi et al. (1995) to calculate radiative 
transfer of molecular lines and to simulate specific molecular line profiles,
respectively. 
For the simulation, we adopted two physical models: Bonnor-Ebert spheres (Bonnor
1956, Ebert 1955), which give only density structure; and Plummer-like spheres 
(Whitworth \& Ward-Thompson 2001), which give density and velocity structures
and can be used for optically thick lines to simulate their self-absorbed 
line profiles (see \S 5.2 of Lee et al. (2003) for details).
We calculated the radiative transfer of dust emission to obtain the temperature 
structure (Evans et al. 2001).
For the structure of molecular abundances, we tested several functional forms;
the step function, which defines a depletion factor ($f_D$) and 
a depletion radius ($r_D$), was the best to fit line profiles from the center
to the outer regions in the three PPCs (Fig. 5 in Lee et al. 2003).
We modeled C$^{18}$O, $\rm H^{13}CO^+$, DCO$^+$, and HCO$^+$ lines using
a step function of each abundance in order to test
how much those molecules are depleted within what radii.
Figure 2 shows that the depletion radius is well constrained around 0.075 pc, 
but the fractional depletion is constrained only to be greater than about 25
in the C$^{18}$O lines of L1512.
This trend occurs in every molecule that is significantly depleted,
so we consider $f_D$ to be the lower limit of the depletion factor.
In other words, the data are consistent with complete depletion inside some
$r_D$.
Figure 3 shows the best-fit model of C$^{18}$O $\rm J=2-1$ in L1512.  
The results of modeling in the three cores are summarized in Table 1.

\begin{figure}
\centering
\includegraphics[width=0.4\linewidth]{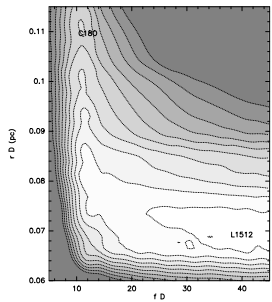}
\caption{The distribution of the reduced $\chi ^2$ of models that have different
$r_D$ and $f_D$ in C$^{18}$O 2$-$1 and 3$-$2 toward L1512.
The contour levels are from 5 to 95 with the interval of 10.
\label{fig-double}}
\end{figure}

\begin{table}
   \begin{center}
    \caption{The results of MC models of molecular lines}\vspace{0em}
    \renewcommand{\arraystretch}{1.1}
    \begin{tabular}[h]{lcccc}
      \hline
       &  &  C$^{18}$O & H$^{13}$CO$^+$ & DCO$^+$ \\
      \hline
     & $X_0$ & $4.82\times10^{-7}$ & $3.0\times10^{-10}$ & $2.8\times10^{-10}$\\
     L1512 & $r_D$  & 0.075 pc & 0.021 pc & $^a$ \\
     & $f_D$ & 25 & 25 & $^a$\\
      \hline
     & $X_0$ & $4.82\times10^{-7}$ & $5.0\times10^{-10}$ & $3.5\times10^{-10}$\\
     L1544 & $r_D$  & 0.045 pc & 0.026 pc & 0.022 pc\\
     & $f_D$ & 25 & 20 & 3 \\
      \hline
     & $X_0$ & $4.82\times10^{-7}$ & $2.2\times10^{-10}$ & $3.5\times10^{-10}$\\
     L1689B & $r_D$ & 0.030 pc & 0.012 pc & 0.011 pc\\
     & $f_D$ & 3 & 5 &4\\
      \hline \\
      \end{tabular}
     \label{tab-single}
   \end{center}
\vskip -0.6cm
{\it\small $^a$ We could not calculate $r_D$ and $f_D$ of DCO$^+$ in L1512
because we have only one spectrum toward the center.}
\end{table}

According to our results, CO is substantially depleted in L1512 and L1544
but not in L1689B. In addition, the depletion radius of CO in L1512 is
greater than that in L1544. The C$^{18}$O lines in L1689B might be affected by
surrounding warm gas (see \S 5.3.2  of Lee et al. 2003).
HCO$^+$ is also depleted significantly in L1512 and L1544, but the depletion 
radii are smaller than those of CO. On the other hand, DCO$^+$ does not show 
significant depletion in any of these three cores. We might not detect the 
depletion of DCO$^+$ because our observational resolution was poor.  

\begin{figure}
\centering
\includegraphics[width=0.9\linewidth]{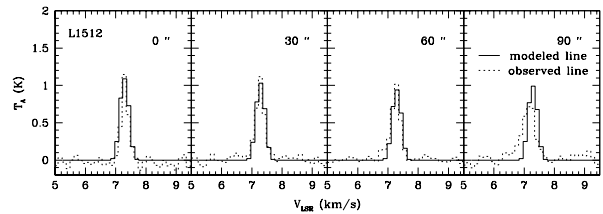}
\caption{The best-fit model of C$^{18}$O $\rm J=2-1$ in L1512. 
The arcseconds marked in panels represent the angular
distance from the dust peak.
\label{fig-double}}
\end{figure}

\section{Discussion}

Our simple analysis and detailed modeling of molecular line profiles combined 
with dust continuum emission show that L1512 is least centrally condensed, but 
CCS, CO, and HCO$^+$ are significantly depleted. Even N$_2$H$^+$ is possibly 
depleted in L1512. 
On the other hand, L1689B has a similar density structure to L1544,
which is quite centrally condensed and has substantial depletion of CCS, CO, and
HCO$^+$, but no molecules except for CCS show significant depletion.
This result may indicate that L1689B has evolved too fast to have time to
evolve chemically.
Therefore, we suggest that the stage of dynamical evolution of a core cannot
be simply probed by its chemical status because chemical evolution depends on
the size of dust grains (Caselli et al. 2002, in preparation) or the absolute 
dynamical timescale, which is different for each dynamical process, as well as 
its density structure that shows the relative dynamical evolutionary stages 
(Table 2). 

\begin{table}
  \begin{center}
    \caption{The results of this study}\vspace{-1.0em}
    \renewcommand{\arraystretch}{1.0}
    \begin{tabular}[h]{lccc}
      \hline
 & L1512 & L1544 & L1689B \\
      \hline
 Density structure &  Young & Evolved & Evolved\\
 Chemical evolution  &  Evolved & Evolved & Young\\
 Timescale   &  $\tau_{che}<<\tau_{dyn}$ &$\tau_{che}\approx\tau_{dyn}$ &
 $\tau_{che}>>\tau_{dyn}$\\
 Dynamical state &Stable (?) &AD (?) & Free-fall (?)\\
      \hline \\
      \end{tabular}
    \label{tab-single}
  \end{center}
\end{table}

\end{document}